\documentclass[
reprint,
twocolumn,
pra,
aps,
superscriptaddress,
nofootinbib
]{revtex4-2}

\usepackage{amsmath,amssymb,amsfonts,bm}
\usepackage{physics}
\usepackage{hyperref}
\usepackage{graphicx}
\usepackage[caption=false]{subfig}

\usepackage{xcolor}

\begin{document}

\title{Quantum Fisher Information and the Speed of Entanglement}

\author{Zain H. Saleem}
\affiliation{Mathematics and Computer Science Division, Argonne National Laboratory, Lemont, Illinois 60439, USA}

\begin{abstract}
We investigate the speed at which entanglement can be generated by an interaction parameter encoded in a two-qubit Hamiltonian, quantified by the derivative of concurrence with respect to the coupling parameter. For arbitrary pure two-qubit states evolving under a general nonlocal interaction, we derive a bound relating this entanglement speed to the quantum Fisher information (QFI). Specifically, we show that
$|\partial_g C| \le \sqrt{F_Q^{(g)}} \sqrt{1-C^2}$,
where $F_Q^{(g)}$ is the QFI associated with estimation of the parameter. This establishes $\sqrt{F_Q}\sqrt{1-C^2}$ as an upper bound on the speed of entanglement generation in parameter space. We further derive the saturation conditions and identify the states and dynamical regimes for which equality is attained. At saturation, concurrence evolves at the maximum rate permitted by the distinguishability of the underlying quantum state. These results reveal a direct connection between quantum metrology and entanglement generation, showing that the same information-theoretic quantity that governs parameter-estimation precision also limits the speed at which entanglement resources can be created.
\end{abstract}

\maketitle

\textit{Introduction.—}
Entanglement is a central resource in quantum information science, underpinning applications ranging from quantum communication and cryptography to quantum computation, quantum networks, and quantum-enhanced sensing \cite{Horodecki2009,Nielsen2010}. For bipartite systems, concurrence occupies a distinguished role as one of the most widely used entanglement measures due to its closed analytical form, direct relation to entanglement of formation, and broad applicability in both theoretical and experimental studies \cite{Wootters1998}. Concurrence has been extensively employed to characterize entanglement generation, degradation under noise, quantum phase transitions, and the performance of entangling gates.

Quantum Fisher information (QFI) is an equally fundamental quantity in quantum metrology \cite{Braunstein1994,Paris2009,Pezze2018}. For a parameter encoded in a quantum state, the QFI determines the ultimate precision achievable by any quantum measurement through the quantum Cramér--Rao bound. Geometrically, QFI quantifies the distinguishability of neighboring quantum states and determines the local statistical speed of evolution in parameter space. Consequently, QFI has become a central tool in quantum sensing, parameter estimation, critical phenomena, and multipartite entanglement detection.

The relationship between entanglement and QFI has traditionally been explored from the perspective of quantum-enhanced metrology, where entanglement acts as a resource capable of increasing the achievable QFI and improving estimation precision \cite{Pezze2018}. Recently, however, attention has shifted toward the inverse question: what constraints does QFI impose on the behavior of entanglement itself? In Ref.~\cite{Saleem2025CoE}, it was shown that the curvature of concurrence with respect to an encoded interaction parameter is bounded by the corresponding QFI, establishing a direct connection between metrological sensitivity and the second-order response of entanglement. Subsequently, Ref.~\cite{saleem2026information} demonstrated that the same quantity governs the leading-order degradation of entanglement under fluctuations of the interaction parameter, providing an information-theoretic bound on the robustness of entanglement generation.

The present work addresses a complementary question. Rather than asking how rapidly the curvature of entanglement changes or how robust entanglement remains under parameter uncertainty, we ask: how rapidly can entanglement itself evolve under variations of an encoded parameter? To answer this question, we consider the first derivative of concurrence with respect to the parameter. We show that the concurrence derivative satisfies $|\partial_g C| \le \sqrt{F_Q^{(g)}} \sqrt{1-C^2}$, where $F_Q^{(g)}$ denotes the quantum Fisher information associated with estimation of the parameter. We further discuss the scope of the two-qubit construction and show that the speed bound admits a natural multiparameter extension in which the QFI is replaced by the corresponding directional form of the QFI matrix.

The result admits a natural geometric interpretation. For pure states, the QFI determines the local statistical distance through the Fubini--Study metric, $ds^2=\frac{1}{4}F_Q^{(g)}dg^2$, so that $\sqrt{F_Q^{(g)}}$ sets the local distinguishability scale between neighboring quantum states. The quantity $\sqrt{F_Q^{(g)}}$ therefore characterizes the maximum rate at which the quantum state itself can change under an infinitesimal variation of the encoded interaction parameter. Our bound shows that the rate at which concurrence changes can never exceed this fundamental distinguishability scale. We further derive the saturation conditions and identify the states and dynamical regimes for which equality is attained. At saturation, concurrence evolves at the maximum rate permitted by the geometry of the underlying quantum state.

These results reveal a direct connection between quantum metrology and entanglement dynamics. While QFI has traditionally been interpreted as a measure of parameter sensitivity and state distinguishability, our results show that it simultaneously constrains the first-order response of an entanglement resource. In this sense, the same information-theoretic quantity that governs the ultimate precision of parameter estimation also determines the maximum speed at which entanglement can evolve. We therefore identify a new role for QFI as a fundamental limit on entanglement dynamics and establish a direct bridge between metrological distinguishability and the speed of entanglement evolution.

\textit{Two-qubit interactions.—}
To make the connection between metrological distinguishability and entanglement dynamics precise, we consider two interacting qubits governed by
\begin{equation}
H(g)=gh,
\end{equation}
where \(g\) denotes the interaction strength and \(h\) is a dimensionless interaction operator. The parameter \(g\) plays a dual role. On the one hand, it controls the entanglement generated by the interaction. On the other, it is the parameter encoded into the evolving quantum state and therefore the quantity whose estimation is governed by the quantum Fisher information. The central question of this work is whether these two notions of sensitivity—the sensitivity of entanglement and the sensitivity of the quantum state itself—are fundamentally related.

The most general two-qubit interaction may be written as
\begin{equation}
h=\sum_{j,k=x,y,z}\eta_{jk}\,
\sigma_j\otimes\sigma_k.
\end{equation}

For the purely nonlocal interaction in Eq.~(2), fixed local basis
transformations leave both concurrence and quantum Fisher information
invariant. The real interaction matrix $\eta$ can therefore be brought,
by singular-value decomposition, to the canonical form
\begin{equation}
h=
\eta_x\,\sigma_x\otimes\sigma_x+
\eta_y\,\sigma_y\otimes\sigma_y+
\eta_z\,\sigma_z\otimes\sigma_z .
\end{equation}
This reduction concerns the purely nonlocal interaction specified in
Eq.~(2); it does not imply that additional local Hamiltonian terms acting
during the evolution can in general be discarded. The canonical representation of nonlocal two-qubit interactions and
their entangling capabilities have been extensively
studied~\cite{Dur2001,Bennett2002}. The canonical form is
particularly useful here because it is diagonal in the Bell basis,
allowing the closed analytical treatment developed below. We later also show that
entanglement-speed bound itself does not require this restriction and
applies to arbitrary differentiable pure two-qubit state families.
The restriction to purely nonlocal interactions is imposed in the
Hamiltonian-based analysis because it permits the explicit Bell-spectrum
treatment and the corresponding Hamiltonian- and initial-state-resolved
characterization developed below.

The canonical interaction is diagonal in the Bell basis, making the Bell states the natural language for describing the dynamics. Expanding an arbitrary pure initial state as
\begin{equation}
|\Psi_0\rangle
=
\sum_{a,b}
\beta_{ab}
|\beta_{ab}\rangle,
\end{equation}
the evolved state becomes
\begin{equation}
|\Psi(g,t)\rangle
=
\sum_{a,b}
\beta_{ab}
e^{-ig\omega_{ab}t}
|\beta_{ab}\rangle,
\end{equation}
where
\begin{equation}
\omega_{ab}
=
(-1)^a\eta_x
-
(-1)^{a+b}\eta_y
+
(-1)^b\eta_z
\end{equation}
are the Bell-state interaction frequencies.

To quantify the entanglement generated by the interaction we employ concurrence, one of the most widely used entanglement measures for two-qubit systems owing to its analytical tractability and direct relation to entanglement of formation \cite{Wootters1998}. For the evolved state, concurrence assumes the form
\begin{equation}
C(g,t)
=
\left|
\sum_{a,b}
(-1)^{a+b}
\beta_{ab}^{\,2}
e^{-2ig\omega_{ab}t}
\right|.
\label{eq:concurrence}
\end{equation}

Equation~(\ref{eq:concurrence}) admits a simple physical interpretation. Each Bell sector contributes a rotating phase factor determined by its interaction frequency. Entanglement generation therefore emerges from interference between different frequency sectors. Constructive interference enhances concurrence, while destructive interference suppresses it. In this picture, concurrence measures the outcome of a frequency-interference process generated by the interaction spectrum.

The sensitivity of the quantum state itself is quantified by the quantum Fisher information. For pure-state evolution generated by Eq.~(1), the QFI with respect to the interaction strength is
\begin{equation}
F_Q^{(g)}
=
4t^2
\left[
\sum_{a,b}
|\beta_{ab}|^2\omega_{ab}^2
-
\left(
\sum_{a,b}
|\beta_{ab}|^2\omega_{ab}
\right)^2
\right].
\label{eq:qfi}
\end{equation}

Unlike concurrence, Eq.~(\ref{eq:qfi}) depends only on the variance of the interaction spectrum sampled by the initial state. Relative phases do not appear. The QFI therefore characterizes the distinguishability of neighboring quantum states, whereas concurrence characterizes the interference responsible for entanglement generation \cite{Braunstein1994,Paris2009,Pezze2018}.

At first sight these quantities appear fundamentally different. One depends on coherent interference between Bell sectors, while the other depends only on spectral variance. Nevertheless, both originate from the same interaction frequencies \(\{\omega_{ab}\}\). This observation naturally raises the question of whether the sensitivity of entanglement is constrained by the sensitivity of the underlying quantum state itself. We now show that the answer is affirmative and that the speed of entanglement evolution is bounded by the square root of the quantum Fisher information.

\textit{Entanglement speed bound and saturation conditions.—}
We now derive the central result of this Letter. To expose the geometric structure underlying the bound, it is convenient to remove the mean interaction frequency
\begin{equation}
\mu=\sum_{a,b}|\beta_{ab}|^2\omega_{ab}
\end{equation}
and define the centered frequencies
\begin{equation}
\Delta_{ab}=\omega_{ab}-\mu.
\end{equation}
Since an overall phase does not affect concurrence, the concurrence amplitude may be written as
\begin{equation}
\widetilde C(g,t)
=
\sum_{a,b}
X_{ab}(g,t),
\qquad
X_{ab}(g,t)
=
(-1)^{a+b}
\beta_{ab}^{\,2}
e^{-2igt\Delta_{ab}},
\end{equation}
with \(C(g,t)=|\widetilde C(g,t)|\). In terms of the same centered frequencies, the QFI takes the variance form
\begin{equation}
F_Q^{(g)}
=
4t^2
\sum_{a,b}
|\beta_{ab}|^2\Delta_{ab}^{\,2}.
\label{eq:centeredqfi}
\end{equation}

We define the entanglement speed in parameter space as
\begin{equation}
v_E(g,t)
=
|\partial_g C(g,t)|,
\end{equation}
which quantifies the first-order response of concurrence to an infinitesimal change in the interaction strength. Because concurrence contains an absolute value, it may fail to be differentiable at isolated points where \(\widetilde C(g,t)=0\). Such points correspond to cusps of the concurrence landscape. In the derivation below we restrict attention to differentiable points. At cusp points, the corresponding one-sided derivatives may be considered separately whenever they exist.

At any differentiable point,
\begin{equation}
|\partial_g C|
=
|\partial_g|\widetilde C||
\le
|\partial_g\widetilde C|.
\label{eq:firstineq}
\end{equation}
Differentiating the concurrence amplitude gives
\begin{equation}
\partial_g\widetilde C
=
-2it
\sum_{a,b}
\Delta_{ab}X_{ab}.
\end{equation}

The centered frequencies satisfy
\begin{equation}
\sum_{a,b}p_{ab}\Delta_{ab}=0,
\qquad
p_{ab}=|\beta_{ab}|^2.
\label{eq:zeromean}
\end{equation}
Defining $z_{ab}=X_{ab}/p_{ab}$ on the support of the state, so that
$|z_{ab}|=1$ and $\widetilde C=\sum_{a,b}p_{ab}z_{ab}$, the derivative
may therefore be written as
\begin{equation}
\partial_g\widetilde C
=
-2it\sum_{a,b}
p_{ab}\Delta_{ab}(z_{ab}-\widetilde C).
\label{eq:centeredderivative}
\end{equation}

The weighted Cauchy--Schwarz inequality then gives
\begin{align}
|\partial_g\widetilde C|^2
&\leq
4t^2
\left(\sum_{a,b}p_{ab}\Delta_{ab}^2\right)
\left(\sum_{a,b}p_{ab}|z_{ab}-\widetilde C|^2\right)
\nonumber\\
&=
F_Q^{(g)}(1-C^2),
\label{eq:CSbound}
\end{align}
where we used
$\sum_{a,b}p_{ab}|z_{ab}-\widetilde C|^2
=1-|\widetilde C|^2=1-C^2$.
Together with Eq.~(\ref{eq:firstineq}), this yields
\begin{equation}
\boxed{
|\partial_g C|
\leq
\sqrt{F_Q^{(g)}}\sqrt{1-C^2}.
}
\label{eq:speedbound}
\end{equation}
The QFI-only relation
$|\partial_gC|\leq\sqrt{F_Q^{(g)}}$ follows immediately.

Equation~(\ref{eq:speedbound}) shows that the QFI sets the
information-geometric scale for entanglement dynamics, while the
instantaneous concurrence further constrains the accessible
entanglement speed. Introducing the normalized speed
\begin{equation}
\nu_C=
\frac{|\partial_gC|}{\sqrt{F_Q^{(g)}}},
\end{equation}
for $F_Q^{(g)}>0$, the bound takes the form
\begin{equation}
\boxed{
C^2+\nu_C^2\leq1.
}
\label{eq:circlebound}
\end{equation}
Thus, the allowed pairs $(C,\nu_C)$ lie within the quarter unit disk,
with saturation corresponding to its circular boundary.

\begin{figure}[t]
    \centering
    \includegraphics[width=0.82\columnwidth]
    {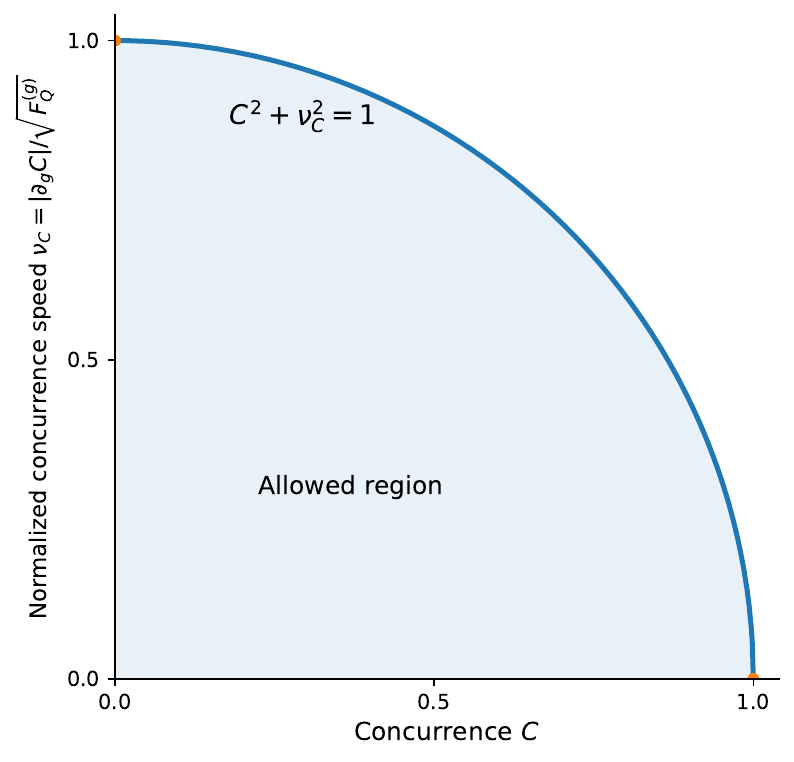}
    \caption{
    Geometric representation of the entanglement-speed bound.
    The concurrence $C$ and normalized concurrence speed
    $\nu_C=|\partial_g C|/\sqrt{F_Q^{(g)}}$ are constrained to the
    quarter unit disk, $C^2+\nu_C^2\leq 1$. The shaded region represents
    the allowed concurrence--speed pairs, while the circular boundary
    $C^2+\nu_C^2=1$ corresponds to saturation of the bound.
    }
    \label{fig:circlebound}
\end{figure}

We now determine the saturation conditions. At a differentiable point
with $0<C<1$ and $F_Q^{(g)}>0$, equality in Eq.~(\ref{eq:firstineq})
requires the phase of $\widetilde C=C e^{i\phi}$ to be stationary,
\begin{equation}
\partial_g\phi=0.
\label{eq:radialcondition}
\end{equation}
Equality in the weighted Cauchy--Schwarz step requires, for every
occupied Bell sector,
\begin{equation}
z_{ab}-\widetilde C=\lambda\Delta_{ab}
\label{eq:CSsaturation}
\end{equation}
for some complex $\lambda$. Combining this condition with
Eqs.~(\ref{eq:centeredderivative}) and (\ref{eq:radialcondition})
gives $\lambda=i\kappa e^{i\phi}$ with $\kappa\in\mathbb{R}$, and hence
\begin{equation}
z_{ab}
=
e^{i\phi}(C+i\kappa\Delta_{ab}).
\label{eq:saturationphasor}
\end{equation}

Since $|z_{ab}|=1$, every occupied sector must satisfy
\begin{equation}
C^2+\kappa^2\Delta_{ab}^2=1,
\end{equation}
and therefore
\begin{equation}
|\Delta_{ab}|=\Delta_*.
\label{eq:twoshell}
\end{equation}
The support is thus confined to the two centered-frequency shells
$\Delta_{ab}=\pm\Delta_*$. 
Defining the total probability carried by the two centered-frequency
shells as
\begin{equation}
P_\pm
=
\sum_{\Delta_{ab}=\pm\Delta_*} p_{ab},
\end{equation}
the zero-mean condition in Eq.~(\ref{eq:zeromean}) gives
\begin{equation}
P_+=P_-=\frac12.
\label{eq:equalweights}
\end{equation}

Writing the two shell phasors as
$z_\pm=e^{i\alpha_\pm}$ and defining
$\delta=\alpha_+-\alpha_-$, their equal weights give
\begin{equation}
\widetilde C
=
\frac{z_++z_-}{2}
=
e^{i(\alpha_++\alpha_-)/2}
\cos\frac{\delta}{2},
\end{equation}
and hence
\begin{equation}
C=\left|\cos\frac{\delta}{2}\right|.
\end{equation}
Similarly, using $\Delta_\pm=\pm\Delta_*$ in
Eq.~(\ref{eq:centeredderivative}) gives
$|\partial_gC|=2t\Delta_*|\sin(\delta/2)|$ under the radiality
condition. Since
$\sqrt{F_Q^{(g)}}=2t\Delta_*$ on the two-shell support,
\begin{equation}
\nu_C=
\frac{|\partial_gC|}{\sqrt{F_Q^{(g)}}}
=
\left|\sin\frac{\delta}{2}\right|.
\end{equation}
Thus the saturation family obeys
$C^2+\nu_C^2=1$.

The occupied Bell sectors must therefore lie on two equally weighted
centered-frequency shells, with their relative phase determining the
concurrence along the saturation boundary. In particular, saturation
is possible at nonzero concurrence and is not restricted to a
particular degree of entanglement.

This structure is closely related to, but distinct from, the saturation of the curvature bound derived in Ref.~\cite{Saleem2025CoE}. In the curvature problem, the second derivative naturally produces quadratic spectral weights \(\Delta_{ab}^{\,2}X_{ab}\), which match the variance appearing in the QFI. 
In the present speed problem, the first derivative instead produces linear
spectral weights $\Delta_{ab}X_{ab}$, and saturation of the resulting
Cauchy--Schwarz bound imposes the additional two-shell condition
$|\Delta_{ab}|=\Delta_*$ on the occupied sectors.
Thus the speed bound characterizes the first-order response of entanglement, while the curvature bound characterizes its second-order response.

An interesting consequence of the saturation conditions is that saturation is not associated with any particular value of the concurrence itself. The conditions involve only the interaction spectrum and the phase structure of the concurrence amplitude. Consequently, product states, partially entangled states, and highly entangled states may all saturate Eq.~(\ref{eq:speedbound}) provided the spectral and phase-alignment conditions are satisfied. The ability of a state to achieve the maximum entanglement speed is therefore determined by the geometry of the interaction dynamics rather than by the amount of entanglement already present.

The speed bound also admits an immediate operational interpretation. For a small perturbation \(\delta g\) of the interaction strength,
\begin{equation}
C(g+\delta g,t)
=
C(g,t)
+
(\partial_g C)\delta g
+
O(\delta g^2).
\end{equation}
Using Eq.~(\ref{eq:speedbound}) therefore gives

\begin{equation}
\left|
\sin^{-1}C(g+\delta g,t)
-
\sin^{-1}C(g,t)
\right|
\leq
\sqrt{F_Q^{(g)}}\,|\delta g|
+
O(\delta g^2).
\label{eq:finitebound}
\end{equation}

Thus the square root of the QFI not only bounds the instantaneous speed of entanglement evolution but also limits the leading-order change in concurrence produced by a small calibration error in the parameter.

\textit{Geometric interpretation and generality.—}
The entanglement-speed bound also admits a direct geometric derivation.
For a pure state $|\psi\rangle$, the geometric measure of entanglement
is defined by its maximal overlap with the set $\mathcal{S}$ of
separable pure states~\cite{Shimony1995,WeiGoldbart2003},
\begin{equation}
E_G(\psi)
=
1-\max_{\phi\in\mathcal{S}}
|\langle\phi|\psi\rangle|^2.
\end{equation}
Equivalently, the Fubini--Study distance from $|\psi\rangle$ to the
separable set is~\cite{Rudnicki2021}
\begin{equation}
D_{\rm FS}(\psi,\mathcal{S})
=
\arcsin\sqrt{E_G(\psi)}.
\end{equation}
For a pure two-qubit state with Schmidt coefficients
$\cos\chi$ and $\sin\chi$, where $0\leq\chi\leq\pi/4$, one has
$E_G=\sin^2\chi$ and $C=\sin 2\chi$ since for two quibits $
E_G=\frac{1-\sqrt{1-C^2}}{2}$. Hence
\begin{equation}
\boxed{
\arcsin C(\psi)
=
2D_{\rm FS}(\psi,\mathcal{S}).
}
\label{eq:concurrenceFS}
\end{equation}

The distance to a fixed set is Lipschitz with respect to the underlying
metric,
\begin{equation}
\left|
D_{\rm FS}(\psi_2,\mathcal{S})
-
D_{\rm FS}(\psi_1,\mathcal{S})
\right|
\leq
D_{\rm FS}(\psi_1,\psi_2).
\label{eq:setdistance}
\end{equation}
For an arbitrary differentiable pure-state family $|\psi(g)\rangle$,
the infinitesimal Fubini--Study line element is
\begin{equation}
ds_{\rm FS}
=
\frac{1}{2}\sqrt{F_Q^{(g)}}\,|dg|.
\end{equation}
Combining this relation with Eqs.~(\ref{eq:concurrenceFS}) and
(\ref{eq:setdistance}) gives
\begin{equation}
\boxed{
\left|
\partial_g\arcsin C
\right|
\leq
\sqrt{F_Q^{(g)}},
}
\label{eq:geometricbound}
\end{equation}
or, equivalently,
\begin{equation}
|\partial_gC|
\leq
\sqrt{1-C^2}\sqrt{F_Q^{(g)}},
\end{equation}
recovering Eq.~(\ref{eq:speedbound}).

This geometric derivation shows that the entanglement-speed bound is a
property of arbitrary differentiable pure two-qubit state families and
does not rely on the canonical nonlocal Hamiltonian or Bell-basis
representation used above. The spectral treatment, however, resolves
the bound in terms of the interaction spectrum and initial-state
amplitudes and provides the explicit saturation conditions. The result
may also be viewed within the broader framework of geometric
entanglement and quantum resource speed limits
~\cite{Rudnicki2021,Campaioli2022}.

\textit{Multiparameter extension:}
The speed bound admits a direct multiparameter formulation. Consider a
pure-state family $|\Psi(\boldsymbol{\theta})\rangle$, with
$\boldsymbol{\theta}=(\theta_1,\ldots,\theta_m)$, and let
$\mathbf{F}_Q$ denote its quantum Fisher information matrix. For an
arbitrary real direction $\mathbf{u}$ in parameter space, define
$\boldsymbol{\theta}(s)=\boldsymbol{\theta}_0+s\mathbf{u}$. The
directional derivative of concurrence is
\begin{equation}
\frac{dC}{ds}=\mathbf{u}^{T}\nabla C,
\end{equation}
while the QFI along this path is
\begin{equation}
F_Q^{(s)}=\mathbf{u}^{T}\mathbf{F}_Q\mathbf{u}.
\end{equation}
Applying Eq.~(19) to this one-dimensional path therefore yields
\begin{equation}
\left|\mathbf{u}^{T}\nabla \arcsin{C}\right|
\leq
\sqrt{\mathbf{u}^{T}\mathbf{F}_Q\mathbf{u}}
\end{equation}

Hence, in the multiparameter setting, the QFI matrix bounds the
concurrence response in every direction of parameter space: the
directional entanglement speed cannot exceed the corresponding
information-geometric speed of the quantum state.

\textit{Discussion and outlook.—}
We have derived a bound relating the speed of entanglement evolution to the quantum Fisher information for arbitrary pure two-qubit states evolving under a general nonlocal interaction. Although the present analysis is restricted to two-qubit pure states, the non-local interaction considered is fully general within this setting, since any two-qubit non-local Hamiltonian can be reduced by local unitaries to the canonical form $( \eta_x X\otimes X + \eta_y Y\otimes Y + \eta_z Z\otimes Z )$, which preserves all nonlocal properties relevant to entanglement generation and QFI. 

The reduction to the canonical form is also closely connected to the theory of the entangling capability of nonlocal two-qubit Hamiltonians. In particular, Dür \textit{et al.} showed that the entanglement-generation capability of a general two-qubit interaction can be characterized through its nonlocal canonical parameters \cite{Dur2001}. Here we use the same general structure for a different purpose: rather than optimizing the entanglement generation rate over states and local controls, we consider the rate of change of entanglement with respect to the interaction parameter, which is directly equivalent to a time derivative under the identification $g  \leftrightarrow t$. 

The bound admits a direct experimental interpretation. Quantum Fisher information has been measured in a variety of platforms including trapped ions, cold atomic ensembles, and superconducting circuits \cite{Strobel2014,Frerot2023}, while concurrence is routinely employed to characterize Bell-state generation, entangling gates, and two-qubit quantum processors \cite{DiCarlo2009,Neeley2010}. Equation~(\ref{eq:speedbound}) therefore provides an experimentally accessible connection between metrological sensitivity and the dynamical response of entanglement.

Our result complements recent bounds on the speed of bipartite entanglement generation based on operator inequalities and Hamiltonian variance \cite{Mohan2023,Mohan2024}. It is also related to recent quantum Fisher information speed limits for multipartite entanglement generation in many-body systems \cite{Chu2023,Chu2024}.  Several directions remain open. It would be interesting to determine whether analogous speed limits exist for mixed-state entanglement measures, multipartite entanglement, and open-system dynamics. Extending the present construction beyond two qubits is nontrivial because there is no unique higher-dimensional analogue of two-qubit concurrence with the same closed analytical structure, while multipartite systems additionally require accounting for inequivalent bipartitions; a systematic treatment of these extensions is therefore left for future work. More generally, the present result raises the possibility that quantum Fisher information may constrain the response of a broader class of quantum resources beyond bipartite entanglement. Establishing such connections could provide a unified information-geometric framework for understanding how quantum resources respond to changes in the parameters that generate them.

\textit{Acknowledgments.—}
The author thanks Anil Shaji and Stephen Gray for helpful discussions over the course of collaborations on related topics. The author also thanks the anonymous referee for providing very valuable suggestions. This work was
supported by Quantum Information Science Enabled
Discovery 2.0 (QuantISED 2.0) for High Energy Physics
(DE-FOA-0003354), with work at Argonne National
Laboratory supported by the U.S. Department of Energy
under Contract No. DE-AC02-06CH11357.

\bibliographystyle{apsrev4-2}
\bibliography{bibliography}

\end{document}